\documentclass[11pt]{article}

\newlength{\medpaperheight}     
\newlength{\medpaperwidth}      
\newlength{\medtextheight}      
\newlength{\medtextwidth}       
\newlength{\medtopmargin}       
\newlength{\medoddsidemargin}   

\newcommand{\real}{{\rm Re}}

\newcommand{\diag}{{\rm diag}}

\renewcommand{\th}{\theta}
\newcommand{\del}{\delta}
\newcommand{\sig}{\sigma}

\newcommand{\Gam}{\Gamma}

\newcommand{\norm}[1]{ \left\| #1 \right\| }

\newcommand{\abs}[1]{\left|#1\right|}

\newcommand{\rhoh}{\hat{\rho}}

\newcommand{\thh}{{\hat{\theta}}}

\newcommand{\fr}[2]{\frac{#1}{#2}}

\newcommand{\eg}{\emph{e.g.}}

\newcommand{\ie}{\emph{i.e.}}

\newcommand{\bquem}{\begin{quote}\begin{em}}
\newcommand{\equem}{\end{em}\end{quote}}

\newcommand{\blist}{\begin{description}}
\newcommand{\elist}{\end{description}}

\newcommand{\bquote}{\begin{quote}}
\newcommand{\equote}{\end{quote}}

\newcommand{\ben}{\begin{enumerate}}
\newcommand{\een}{\end{enumerate}}

\newcommand{\bit}{\begin{itemize}}
\newcommand{\eit}{\end{itemize}}

\newcommand{\bea}{\begin{array}}
\newcommand{\eea}{\end{array}}

\newcommand{\bds}{\begin{displaystyle}}
\newcommand{\eds}{\end{displaystyle}}

\newcommand{\Rbf}{{\mathbf R}}
\newcommand{\Cbf}{{\mathbf C}}

\newcommand{\ds}{\displaystyle}

\newcommand{\ns}{\normalsize}

\newcommand{\refeq}[1]{(\ref{eq:#1})}

\newcommand{\set}[2]{ \left\{ \,#1\, \left| \,#2\, \right.\right\} }

\newcommand{\seq}[1]{ \left\{ #1 \right\} }

\newcommand{\sett}[2]{ \Big\{ \,#1\, \Big| \,#2\, \Big\} }

\makeatletter
\def\beq{\@ifnextchar 
[{\@tempswatrue\@beq}{\@tempswafalse\@beq[]}}
\def\@beq[#1]{\begin{equation}\edef\@tmparg{#1}\ifx\@tmparg\@e
mpty \else
	\label{#1}\fi}
\newcommand{\eeq}{\end{equation}}
\makeatother 
\newcommand{\beqaa}{\begin{eqnarray*}}
\newcommand{\eeqaa}{\end{eqnarray*}}

\newcommand{\beqa}{\begin{eqnarray}}
\newcommand{\eeqa}{\end{eqnarray}}

\newcommand{\bc}{\begin{center}}
\newcommand{\ec}{\end{center}}

\usepackage{epsfig}
\usepackage{times}
\usepackage{amsmath,amssymb}

\newcommand{\fro}{{\rm F}}
\newcommand{\Ebf}{{\bf E}}
\newcommand{\Phih}{\hat{\Phi}}
\newcommand{\Phib}{\bar{\Phi}}

\newcommand{\fpur}{f}
\newcommand{\fmix}{\underline{f}}
\newcommand{\favg}{\overline{f}}

\newcommand{\Scal}{ {\cal S} }

\renewcommand{\ns}{{n_S}}
\newcommand{\nb}{{n_B}}
\newcommand{\Cbfss}{\Cbf^{\ns\times\ns}}
\newcommand{\Cbfbb}{\Cbf^{\nb\times\nb}}
\newcommand{\Cbfsbsb}{\Cbf^{\ns\nb\times\ns\nb}}

\renewcommand{\sett}[2]{ \left\{ \,#1\, \Big| \,#2\, \right\} }

\newcommand{\Usb}{U}
\newcommand{\Udes}{G}
\newcommand{\Ub}{U_B}
\newcommand{\Us}{U_S}

\renewcommand{\norm}[1]{ \left\| #1 \right\| }
\renewcommand{\bar}[1]{\overline{#1}}

\renewcommand{\hat}[1]{\widehat{#1}}

\newcommand{\normf}[1]{\norm{#1}_{\fro} }

\newcommand{\normtr}[1]{\norm{#1}_{\rm Tr}}

\newcommand{\Hcal}{{\cal H}}

\renewcommand{\rhoh}{\hat{\rho}}

\newcommand{\avg}[1]{\Ebf\left\{#1\right\}}
\newcommand{\trace}{{\bf Tr}}

\newcommand{\ket}[1]{| #1 \rangle}
\newcommand{\bra}[1]{\langle #1 |}
\newcommand{\braket}[2]{\langle #1 | #2 \rangle}

\renewcommand{\hbar}{ {h {\!\!\!^{\scriptscriptstyle -} } } }

\newcommand{\prob}[1]{ {\bf Prob}\left\{ #1 \right\} }

\begin{document}

\title{
On the distance between unitary propagators of quantum systems
of differing dimensions 
}
\author{
Robert L. Kosut
\\
SC Solutions, Sunnyvale, CA
\\
{\tt kosut@scsolutions.com}
\and
Matthew Grace, Constantin Brif, Herschel Rabitz
\\
Department of Chemistry
\\
Princeton University, Princeton, NJ
%\and
%Constantin Brif\\ Department of Chemistry\\ 
%Princeton University, Princeton, NJ
}

\date{}

\maketitle
%%%%%%%%%%%%%%%%

%%%%%%%%%%%%%%%%%%%%%%%%%%%%%%%%%
\begin{abstract}

\noindent
A distance measure is presented between two unitary propagators of
quantum systems of differing dimensions along with a corresponding
method of computation. A typical application is to compare the
propagator of the actual (real) process with the propagator of the
desired (ideal) process; the former being of a higher dimension then
the latter. The proposed measure has the advantage of dealing with
possibly correlated inputs, but at the expense of working on the whole
space and not just the information bearing part as is usually the
case, \ie, no partial trace operation is explicitly involved.  It is
also shown that the distance measure and an average measure of channel
fidelity both depend on the size of the same matrix: as the matrix
size increases, distance decreases and fidelity increases.

%A measure is presented which quantifies the distance between the
%propagators of two quantum systems with differing dimensions.
\end{abstract}

%%%%%%%%%%%%%%%%%%%%%
\section{Introduction}

Much of engineering design is predicated on having a model of the
process which captures the essential features (\eg, the design
variables and sources of uncertainty), a model of the desired process,
and a means of comparing the two. In addition, it is also desirable
that the same means of comparison can be determined or validated from
experimental data.  Distance and fidelity measures are proposed in
\cite{GilchristLN:04} for comparing what are referred to as real and
ideal quantum processes represented as {\em quantum operations}
\cite{NielsenC:00}. Several measures are proposed in
\cite{GilchristLN:04} along with a set of criteria which aim at making
the measure tractable both theoretically as well as
experimentally. Here we focus more on the theoretical aspect and in
particular using unitary propagators corresponding to the real and
ideal systems.  The proposed distance measure does, however, have an
input output interpretation, and thus, in principal, it may be
amenable to an experimental calculation although that is not pursued
here.

%%%%%%%%%%%%%%%%%%%%%%%%%%%%%%%%
\section{Problem formulation}

Consider a finite dimensional, closed, bipartite quantum system
consisting of a part, $S$, referred to as the {\em system}, which
bears the quantum information of interest, coupled to a part $B$,
representing the {\em bath} or {\em environment}.  The Hilbert space
of the bipartite system is $\Hcal_{SB}=\Hcal_S\otimes\Hcal_B$ where
$\Hcal_S$ and $\Hcal_B$ are finite dimensional with $\ns=\dim\Hcal_S$,
$\nb=\dim\Hcal_B$, and hence, $\ns\nb=\dim\Hcal_{SB}$. Let
$\Usb\in\Cbfsbsb$ denote the unitary propagator of the (real)
bipartite system, and let $\Udes\in\Cbfss$ denote the unitary
propagator of an ideal system acting only on $\Hcal_S$. The question
addressed here is this:
\bquote 
How close is $\Usb$ to $\Udes$?  
\equote
Since it is implicitly assumed that experimental access is confined to
the $S$-system, it follows that performance is judged only by how well
the reduced state output mimics the desired behavior in $\Hcal_S$ as
represented by $\Udes$. If, for example, $\Usb$ decomposes into a
tensor product of unitaries, \ie, $U=\Udes\otimes\Ub$, then the $S$
and $B$ systems do not interact and the system is indistinguishable
from $\Udes$ acting on $\Hcal_S$.  This motivates expressing $\Usb$
as,\footnote{
If the $S$ and $B$ channels are ordered reversely, then
$\Usb=\Phi\otimes\Udes+R$.
}
\beq[eq:usb]
\Usb=\Udes\otimes\Phi+R
\eeq
where $\Phi\in\Cbfbb$ is unitary. Since any choice of $\Phi$
determines $R$, it follows that it is always possible to express
$\Usb$ in this form. Hence, consider the following distance measure
between $\Usb\in\Cbfsbsb$ and $\Udes\in\Cbfss$.
\beq[eq:dist]
d(\Usb,\Udes)
=
\min_\Phi
\sett{ \norm{\Usb-\Udes\otimes\Phi} }
{ \Phi^\dag\Phi=I_B }
\eeq
where $\norm{\cdot}$ is any matrix norm on $\Cbfsbsb$.
%
%and $\alf$ is a scaling factor which normalizes the distance
%measure to the range $[0,1]$.
%
The term inside the norm is $R$ as defined in \refeq{usb}.  The
unitary $\Phi\in\Cbfbb$ is thus used solely to find the closest
propagator to $\Usb$ in the subspace
$\sett{\Udes\otimes\Phi}{\Phi^\dag\Phi=I_B}$ where ``closeness'' is
determined by the choice of norm.

It is clear that using \refeq{dist} as a distance measure does not
consider the partial trace operation which is implicit in other
distance (and fidelity) measures via the operator-sum-representation
(OSR), \eg, \cite{NielsenC:00}, \cite{KretschmannW:04},
\cite{GilchristLN:04}. As a result \refeq{dist} is in general a
conservative measure. More specifically, \refeq{dist} is measuring
distance over the whole space $\Hcal_S\times\Hcal_B$ rather than just
over $\Hcal_S$. For example, suppose the (possibly correlated) pure
state $\ket{\psi}$ is acted upon by $\Usb$ and its output is compared
to the ideal output of $\Udes\otimes\Phi$ acting on $\ket{\psi}$.
The output state error is,
\beq[eq:kete]
e=(\Usb-\Udes\otimes\Phi)\ket{\psi}
\eeq
It follows that 
\beq[eq:sv]
\max_{\braket{\psi}{\psi}=1}\
e^\dag e
=
\norm{\Usb-\Udes\otimes\Phi}_2^2
\eeq
where $\norm{\cdot}_2$ is the induced two-norm, \ie, the maximum
singular value of the matrix argument.  Suppose the input is random in
the sense that $\ket{\psi}=\ket{\psi_i}$ with probability $p_i$ for
$i=1,\ldots,\ell$. Let $\Ebf$ denote the expected value operator taken
with respect to the input distribution. Then,
\beq[eq:fro]
\bea{rcl}
\ds
\max_{
\prob{\ket{\psi}=\ket{\psi_i}}=p_i
}
\Ebf\
e^\dag e
&=&
\ds
\max_{
\rho=\sum_i p_i\ket{\psi_i}\bra{\psi_i}
}
\trace\ \left(\Usb-\Udes\otimes\Phi\right)^\dag
\left(\Usb-\Udes\otimes\Phi\right)\ \rho
\\
&=&
\ds
\max_{
\rho\geq 0,\ \trace\ \rho=1
}
\trace\ \left(\Usb-\Udes\otimes\Phi\right)^\dag
\left(\Usb-\Udes\otimes\Phi\right)\ \rho
\\
&=&
\ds
\normf{\Usb-\Udes\otimes\Phi}^2/\ns\nb
\eea
\eeq
where $\normf{\cdot}$ is the Frobenius matrix norm:
$
\normf{X}
=
\left(\trace\ X^\dag X\right)^{1/2}
=
\left(\sum_{i,j}\abs{X_{ij}}^2\right)^{1/2}
=
\left(\sum_{i}\sig_i(X)^2\right)^{1/2}
$
with $\sig_i(X)$ the $i$-th singular value of $X$. 
%
\iffalse
Recall that the two matrix norms above satisfy,
%
$
\norm{\Usb-\Udes\otimes\Phi}_2
\leq
\normf{\Usb-\Udes\otimes\Phi}
\leq
\sqrt{\ns\nb} \norm{\Usb-\Udes\otimes\Phi}_2
$.
\fi
%
Since the error is defined over states operating on the whole of
$\Hcal_S\times\Hcal_B$, \refeq{sv}-\refeq{fro} are upper bounds on the
error when the actual input is uncorrelated, that is, when either
$\ket{\psi}=\ket{\psi_S}\otimes\ket{\psi_B}$ or
$\rho=\rho_S\otimes\rho_B$. Despite this, however, there is a benefit:
using either the maximum singular value or the Frobenius norm in
\refeq{dist} gives an indication of {\em both} the effect of
environmental input uncertainty (including unwanted correlations with
the information input) and the ``distance'' of the actual system to an
ideal. In the remainder of this note we will show how to compute the
distance measure using both of the above matrix norms. We also explore
the relation of this distance measure to a typical fidelity measure.

%%%%%%%%%%%%%%%%%%%%%%
\section{Computing the distance measure with the Frobenius norm}

Consider the distance measure
\beq[eq:dfro]
d_\fro(\Usb,\Udes)
=
\min_\Phi
\sett{ \frac{1}{\sqrt{2\ns\nb}}\normf{\Usb-\Udes\otimes\Phi} }
{ \Phi^\dag\Phi=I_B }
\eeq
where $1/\sqrt{2\ns\nb}$ is a normalization factor whick keeps
$d_\fro(\Usb,\Udes)$ in the range $[0,1]$.  As shown below,
\beq[eq:distf]
\bea{rcl}
\ds
d_\fro(\Usb,\Udes)
&=&
\ds
\left(
1 - \fr{1}{\ns\nb} \normtr{\Gam}
\right)^{1/2}
\\&&\\
\Gam
&=&
\ds
\sum_{i,j=1}^\ns \Udes_{ij}^* \Usb_{(ij)}
\in\Cbfbb
\eea
\eeq
where 
$\normtr{\cdot}$ is the matrix {\em trace-norm},\footnote{
For $X\in\Cbf^{n\times m}$, $\normtr{X}=\trace\ \sqrt{X^\dag X} =
\sum_{i=1}^{\min(n,m)}\ \sig_i(X)$ where $\sig_i(X)$ is the $i$-th
singular value of $X$.
}
the $\seq{\Usb_{(ij)}}$ are $\nb\times\nb$ matrix partitions of
$\Usb$,
\beq[eq:usbij]
\Usb
=
\left[
\bea{ccc}
\Usb_{(11)} & \cdots & \Usb_{(1\ns)}
\\
\vdots & \vdots & \vdots
\\
\Usb_{(\ns 1)} & \cdots & \Usb_{(\ns\ns)}
\eea
\right]
\in\Cbfsbsb
\eeq
and $\seq{\Udes_{ij}}$ are the (scalar) elements of $\Udes$,
\beq[eq:udesij]
\Udes
=
\left[
\bea{ccc}
\Udes_{11} & \cdots & \Udes_{1\ns}
\\
\vdots & \vdots & \vdots
\\
\Udes_{\ns 1} & \cdots & \Udes_{\ns\ns}
\eea
\right]
\in\Cbfss
\eeq
%

%%%%%%%%%%%%%%%%%%%%%%%%%%%
\bquote
{\bf Proof}

From the definition \refeq{dfro},
\beq
\bea{rcl}
\ds
\frac{1}{2\ns\nb}\normf{\Usb-\Udes\otimes\Phi}^2
&=&
\ds
1-\frac{1}{\ns\nb}\
\real\ \trace\ (\Udes^\dag\otimes\Phi^\dag)\Usb
\\&&\\
&=&
\ds
1-\frac{1}{\ns\nb}\
\real\ \trace\
\sum_{i,j=1}^\ns\
\Udes_{ij}^*\ \Usb_{(ij)}
\Phi^\dag
\\&&\\
&=&
\ds
1-\frac{1}{\ns\nb}\
\real\ \trace\ \Gam\Phi^\dag
\eea
\eeq
The first line above follows from the fact that $\Usb,\ \Udes,\ \Phi$
are all unitary.  The last two lines follow from the definition of the
tensor product together with \refeq{distf}-\refeq{udesij}. To compute
$d_\fro(\Usb,\Udes)$ is equivalent to finding the maximum value of
$\real\ \trace\ \Gam\Phi^\dag$ over all unitary $\Phi\in\Cbfbb$. A
singular value decomposition of $\Gam$ gives $\Gam=WSV^\dag$ with
unitary $W$ and $V$ and with the singular values in the diagonal
matrix $S=\diag(\sig_1,\ldots,\sig_\nb)\in\Rbf^\nb,\ \sig_1 \geq
\sig_2 \geq \cdots \geq \sig_\nb \geq 0$. This gives,
$
\real\ \trace\ \Gam\Phi^\dag 
= 
\real\ \trace\ WSV^\dag\Phi^\dag 
= \real\ \trace\ SV^\dag\Phi^\dag W 
= \real\ \trace\ SB
$
with $B=V^\dag\Phi^\dag W$. Clearly $\Phi$ is unitary if and only if
$B$ is unitary. Hence, $\real\ \trace\ SB = \sum_{i=1}^\nb\ s_i\
\real\ B_{ii}$ will achieve the maximum, $\real\ \trace\ S=\trace\ S
=\normtr{\Gam}$ when $\real\ B=I$, if and only if $B=I$, which is of
course unitary. This completes the proof of \refeq{distf}.

\equote
%%%%%%%%%%%%%%%%%%

%%%%%%%%%%%%%%%%%%%%%%%%%%%%%%
\subsection*{Reversing the channel ordering}

By an analogous argument, if the $S$ and $B$ channels are ordered
reversely from \refeq{usb}, then the distance measure becomes,
\beq[eq:distfrev]
\bea{rcl}
\ds
d_\fro(\Usb,\Udes)
&=&
\ds
\min_\Phi
\sett{ \frac{1}{\sqrt{2\ns\nb}}\normf{\Usb-\Phi\otimes\Udes} }
{ \Phi^\dag\Phi=I_B }
\\&&\\
&=&
\ds
\left(
1 - \fr{1}{\ns\nb}\normtr{\Gam}
\right)^{1/2}
\\&&\\
\Gam_{ij} 
&=&
\ds
\trace\ \Udes^\dag \Usb_{(ij)},\
i,j=1,\ldots,\nb
\eea
\eeq
where now $\seq{\Usb_{(ij)}}$ are $\ns\times\ns$ matrix partitions of $\Usb$,
\beq[eq:usbijrev]
\Usb
=
\left[
\bea{ccc}
\Usb_{(11)} & \cdots & \Usb_{(1\nb)}
\\
\vdots & \vdots & \vdots
\\
\Usb_{(\nb 1)} & \cdots & \Usb_{(\nb\nb)}
\eea
\right]
\in\Cbfsbsb
\eeq
%

%%%%%%%%%%%%%%%%%%%%%%%%%%%%
\subsection*{Exact tensor product}

Suppose $\Usb$ is exactly a tensor product, \ie,
\beq[eq:exact]
\Usb = \Us \otimes \Ub,
\eeq
Then, 
\beq[eq:dexact]
\bea{rcl}
U_{(ij)} &=& (\Us)_{ij} \Ub
\\&&\\
\Gam
&=&
\sum_{i,j=1}^\ns \Udes^*_{ij}(\Us)_{ij}\Ub
=
\left( \trace\ \Udes^\dag\Us \right)\Ub
\\&&\\
\normtr{\Gam}
&=&
\trace\sqrt{\abs{\trace\ \Udes^\dag\Us}^2 \Ub\Ub^\dag}
=
\abs{\trace\ \Udes^\dag\Us}\trace I_B
=
\nb\abs{\trace\ \Udes^\dag\Us}
\eea
\eeq
The distance measure becomes,
\beq[eq:dexact01]
d_\fro(\Us\otimes\Ub,\Udes)
=
\left(
1-\fr{1}{\ns}\abs{\trace\ \Udes^\dag\Us}
\right)^{1/2}
\eeq
If $\Usb$ and $\Udes$ have the same dimensions, that is, $\nb=1$, then
the distance is zero if and only if $\Usb\ (\equiv \Us)$ and $\Udes$
differ by a scalar phase, \ie, $\Us = e^{i\phi}\Udes$ for some real
phase angle $\phi$.

%%%%%%%%%%%%%%%%%%%%%%%%%%%%%
\section{Computing the distance measure with the maximum singular value}

Consider the distance measure
\beq[eq:d2]
d_2(\Usb,\Udes)
=
\min_\Phi
\sett{ \frac{1}{\sqrt{2}}\norm{\Usb-\Udes\otimes\Phi}_2 }
{ \Phi^\dag\Phi=I_B }
\eeq
where $1/\sqrt{2}$ is a normalization factor which keeps
$d_2(\Usb,\Udes)$ in the range $[0,1]$. In this case we have not been
able to obtain an exact solution, but we can establish the following
bounds:
\beq[eq:d2bnd]
\frac{1}{\sqrt{2}}
\norm{\Usb-\Udes\otimes\Phib}_2
\leq
d_2(\Usb,\Udes)
\leq
\frac{1}{\sqrt{2}}
\norm{\Usb-\Udes\otimes\Phih}_2
\eeq
where $\Phib$ is the solution to the convex optimization problem:
\beq[eq:phib]
\bea{ll}
\mbox{minimize}
&
\norm{\Usb-\Udes\otimes\Phi}_2
%\\&\\
\\
\mbox{subject to}
&
\Phi^\dag\Phi \leq I_B
\eea
\eeq
and where $\Phih$ is obtained from $\Phib$ via the singular value
decomposition of $\Phib$ as follows:
\beq[eq:phih]
\Phib=VSW^\dag\ \Rightarrow\ \Phih=VW^\dag
\eeq
The upper and lower bounds in \refeq{d2bnd} will be close to each
other if the singular values of $\Phib$ are close to unity, that is,
if $\Phib$ is close to a unitary.

%%%%%%%%%%%%%%%%%%%%%%%
\bquote
{\bf Proof}

Problem \refeq{phib} is a convex optimization because any norm is a
convex function, the argument in the norm is affine in the
optimization variable $\Phi$, and the constraint set $\Phi^\dag\Phi
\leq I_B$ is convex \cite{BoydV:04}. However, there is no guaranty
that the resulting optimizer $\Phib$ is unitary, \ie, on the boundary
of the constraint. Since $\Phi$ in \refeq{phib} is less constrained
then in \refeq{d2}, it follows that the lower bound in \refeq{d2bnd}
applies. In consequence, consider $\Phih$ from \refeq{phih} as a
unitary approximation to $\Phib$, and since it is not necessarily the
optimal solution to \refeq{d2}, the upper bound in \refeq{d2bnd}
follows which completes the proof.

\equote
%%%%%%%%%%%%%%%%%%%%%%%%%%%

%%%%%%%%%%%%%%%%%%%%
\section{Application to optimal control or system design}

In the ideal case where $\Usb$ and $\Udes$ are of the same dimension,
the term $\abs{\trace\ \Udes^\dag\Us}$ in \refeq{dexact01} has often
been proposed as an objective function to be maximized for optimal
control design (\eg, \cite{PalaoK:02}) and also for determining
properties of the control landscape \cite{RabitzHR:04}. This is
equivalent to posing the distance measure \refeq{dexact01} as an
objective function to be minimized.  Where $\Usb$ and $\Udes$ are not
of the same dimension, the distance measure \refeq{distf} has been
reported in \cite{Grace:06aps} for optimal control design.

In general, for either control design or system design, $\Usb$ will
depend on some parameters. In many cases these parameters are
constants, \eg, coefficients of specified time functions which make up
the control field, settings of wave-plate angles in a photonic device,
geometry variables in a circuit layout, and so on. In these cases,
$\Usb\equiv\Usb(\th)$ where $\th$ is a (constant) vector to be
selected out of a set $\Theta$ to minimize the distance measure
\refeq{dist}. Equivalently, consider the following optimization
problem:
\beq[eq:opt]
\bea{ll}
\mbox{minimize}
&
%\mbox{$d_\fro(\Usb(\th),\Udes)$ or $d_2(\Usb(\th),\Udes)$}
\norm{\Usb(\th)-\Udes\otimes\Phi}
\\
\mbox{subject to}
&
\Phi^\dag\Phi=I_B,
\;\;
\th\in\Theta
\eea
\eeq
where $\norm{\cdot}$ is any matrix norm and the optimization variables
are $\Phi$ and $\th$. Although typically $\Theta$ is a convex set,
\eg, $\Theta=\set{\th}{\norm{\th-\th_0}\leq\del}$, in general
\refeq{opt} is not a convex optimization. In the first place, it is
almost never the case that $\Usb(\th)$ is a convex function of
$\th$. Suppose for example, that the Hamiltonian which generates
$\Usb(\th)$ depends linearly on $\th$, \ie, $H(\th,t)=\sum_k\ \th_k\
H_k(t)$. If $\Usb(\th)$ is the associated propagator which makes its
appearance at some time $\tau$, then it is unlikely that $\Usb(\th)$
is convex over $\Theta$. Secondly, the constraint $\Phi^\dag\Phi=I_B$
is not a convex set. This constraint, however, can be relaxed to the
convex set $\Phi^\dag\Phi\leq I_B$ resulting in,
\beq[eq:optrlx]
\bea{ll}
\mbox{minimize}
&
%\mbox{$d_\fro(\Usb(\th),\Udes)$ or $d_2(\Usb(\th),\Udes)$}
\norm{\Usb(\th)-\Udes\otimes\Phi}
\\
\mbox{subject to}
&
\Phi^\dag\Phi\leq I_B,
\;\;
\th\in\Theta
\eea
\eeq
Since $\norm{\Usb(\th)-\Udes\otimes\Phi}$ for any norm is a convex
function of $\Phi$ and $\Usb(\th)$, and $\Phi^\dag\Phi\leq I_B$ is a
convex set in $\Phi$, it follows that \refeq{optrlx} is a convex
optimization over $\Usb(\th)$ and $\Phi$ (see, \eg, \cite[\S
3.25]{BoydV:04}).  Again, this is not a convex optimization over $\th$
because $\Usb(\th)$ is not a convex function of $\th$.  Nevertheless,
the following iterative scheme will always find a local solution by
reducing the distance measure in every step.
\bquote
\begin{tabular}{ll}
{\bf Initialize}
&
$\thh=\th_0$
\\
{\bf Repeat}
&
1. 
$
\ds
\Phih
=
\arg\min_\Phi\sett{\norm{\Usb(\thh)-\Udes\otimes\Phi}}{\Phi^\dag\Phi=I_B}
$
\\
&
2.
$
\ds
\thh
=
\arg\min_\th\sett{\norm{\Usb(\th)-\Udes\otimes\Phih}}{\th\in\Theta}
$
\\
{\bf Until} 
&
$\norm{\Usb(\thh)-\Udes\otimes\Phih}$ stops decreasing.
\end{tabular}
\equote
In step 1 if the Frobenius norm is used then $\Phih$ can be calculated
exactly \refeq{distf}. Under the two-norm we would use the
approximation from \refeq{phih}.  Step 2 requires using a local
solver.

%%%%%%%%%%%%%%%%%%%%%%%%
\section{Distance and fidelity}

We will show that the distance measure defined here and a typical
measure of fidelity both depend on the size of the matrix $\Gam$ as
given by either \refeq{distf} or \refeq{distfrev}.  There are many
ways that fidelity has been defined to compare the desired unitary
$\Udes$ with a quantum channel, \eg, \cite{NielsenC:00},
\cite{KretschmannW:04}, \cite{GilchristLN:04}.  Specifically, let
$\Scal$ denote a trace-preserving quantum channel mapping states
$\rho_S\in\Hcal_S$ to $\rhoh_S\in\Hcal_S$ with the
operator-sum-representation (OSR),
\beq[eq:srho] 
\rhoh_S=\Scal(\rho_S) = \sum_k S_k \rho_S S_k^\dag,     
\;\;\;
\sum_k S_k^\dag S_k = I_S
\eeq
with (matrix) operation elements $S_k\in\Cbfss$.  Consider, for
example, the {\em worst-case pure state fidelity},
\beq[eq:fpure]
\fpur(\Scal,\Udes) 
=
\min_{\ket{\psi_S}}\
(\Udes\ket{\psi_S})^\dag \rhoh_S (\Udes\ket{\psi_S}) 
=
\min_{\ket{\psi_S}}  
\sum_k | \bra{\psi_S} \Udes^\dag S_k \ket{\psi_S} |^2
\eeq
where
$
\rhoh_S = \sum_k\ S_k \ket{\psi_S}\bra{\psi_S} S_k^\dag
$
is the reduced output state of $\Scal$ with $\ket{\psi_S}$ the pure
input state.  In general calculating $\fpur(\Scal,\Udes)$ is not easy
as it is not a convex optimization; it is, however, bounded as
follows:
\beq[eq:fid ineq] 
\fmix(\Scal,\Udes) \leq \fpur(\Scal,\Udes) \leq \favg(\Scal,\Udes)
\eeq
where
\beq[eq:fid def]
\bea{rcl}
\favg(\Scal,\Udes) 
&=& 
\ds
\frac{1}{\ns^2} \sum_k | \trace \Udes^\dag S_k  |^2
\\&&\\
\fmix(\Scal,\Udes) 
&=&
\ds 
\min_{\rho_S} \sum_k | \trace\ \Udes^\dag S_k \rho_S |^2
\eea
\eeq
All these fidelities are in [0,1] and equal to one if and only if
$\Scal(\rho_S)=\Udes\rho_S\Udes^\dag$ for all densities $\rho_S$.
%
%(Equivalently, as shown in \cite{KnillL:97}, if there exist scalars
%$\seq{\alf_k}$ such that $S_k=\alf_k\Udes,\ \sum_k|\alf_k|^2=1$.)
%
Both bounds are easy to obtain: $\favg(\Scal,\Udes)$ is direct and
$\fmix(\Scal,\Udes)$ is a convex optimization over all densities (\ie,
over all $\rho_S\in\Cbfss,\ \rho_S\geq 0,\ \trace\ \rho_S=1$), and
hence, can be numerically obtained. If the worst-case density
associated with $\fmix(\Scal,\Udes)$ is nearly rank one, then
$\fmix(\Scal,\Udes) \approx \fpur(\Scal,\Udes)$.

In order to relate $\Usb$ to $\Scal$ assume that the input state to
$\Usb$ is the pure uncorrelated state\footnote{
For convenience we use what we referred to previously as the {\em
reversed} channel ordering.
}
$\ket{\psi_B}\otimes\ket{\psi_S}$ where $\ket{\psi_B}$ has elements
$\psi_{Bi},\ i=1,\ldots,\nb$ with $\sum_i\abs{\psi_{Bi}}^2=1$. It then
follows that
\beq[eq:u2s]
S_k = \sum_{i=1}^\nb\ \psi_{Bi} \Usb_{(ki)},\
k=1,\ldots,\nb
\eeq
where the $\Usb_{(ki)}$ are the $\ns\times\ns$ matrix partitions of
$\Usb$ as given by \refeq{usbijrev}. The upper bound in \refeq{fid
ineq} is then,
\beq[eq:fup]
\favg(\Usb,\Udes)
=
\frac{1}{\ns^2}
\sum_{i,j=1}^\nb\
\psi_{Bi}^{}\psi_{Bj}^*
\left(
\sum_{k=1}^\nb\
\Gam_{ki}\Gam_{kj}^*
\right)
=
\frac{1}{\ns^2}
\bra{\psi_B} \Gam^\dag\Gam\ket{\psi_B}
\eeq
with $\Gam_{ki}=\trace\ G^\dag U_{(ki)}$ from \refeq{distfrev}.  This
upper bound on pure state fidelity thus depends on $\Gam$ weighted by
the state of the environment $\ket{\psi_B}$.  Suppose the environment
state is completely random, that is,
$\avg{\ket{\psi_B}\bra{\psi_B}}=I_B/\nb$.  Then the average value of
the pure state fidelity upper bound becomes,
\beq[eq:fupavg]
\avg{\favg(\Usb,\Udes)}
=
\frac{1}{\ns^2\nb}
\sum_{k,i=1}^\nb\
\abs{\Gam_{ki}}^2
=
\frac{1}{\ns^2\nb}
\normf{\Gam}^2
\eeq
The same result holds with $\Gam$ defined by \refeq{distf}. Thus the
distance measure presented here and a typical channel fidelity both
depend on the matrix $\Gam$ defined by either \refeq{distf} or
\refeq{distfrev}, depending on how channel ordering is
ascribed. Increasing the size of $\Gam$ increases fidelity and
decreases distance.

We also mention that the inequality relation of distance and fidelity
described in \cite{GilchristLN:04} also holds here as well,
namely,
\beq[eq:fiddist01]
\left(1-d_\fro(\Usb,\Udes)\right)^2 \leq \avg{\favg(\Usb,\Udes)} \leq
1-d_\fro(\Usb,\Udes)^2 
\eeq

\begin{quote}
{\bf Proof}

We first show that the bounds follow from
$\norm{\Gam}_2=\max_i\sig_i\leq\ns$ where $\sig_i$ is the $i$-th
singular value of $\Gam$. Then we will show that
$\norm{\Gam}_2\leq\ns$ holds.

The upper bound holds if and only if $\normf{\Gam}^2 \leq \ns
\normtr{\Gam}$, or equivalently, $\sum_i\sig_i^2 \leq \ns\sum_i
\sig_i$. Assuming that $\sig_i\leq\ns$, it immediately follows that
$\sum_i\sig_i^2 \leq \ns\sum_i \sig_i$ which establishes the upper
bound.

The lower bound is equivalent to $\normtr{\Gam}+(1/\ns)\normf{\Gam}^2
\leq 2\sqrt{\nb}\normf{\Gam}$. Assuming again that
$\norm{\Gam}_2\leq\ns$, and hence, $\normf{\Gam}^2 \leq \ns
\normtr{\Gam}$, the lower bound will hold if
$\normtr{\Gam}\leq\sqrt{\nb}\normf{\Gam}$. This is a known inequality
for any matrix and thus establishes the lower bound.

To show that $\sig_i\leq\ns$, observe that since the elements of
$\Gam$ \refeq{distfrev} are obtained via the trace operation, then by
definition they are equivalently obtained from the sum of the diagonal
elements of $\Udes^\dag\Usb_{(ij)}$. Let $U'$ denote the matrix whose
$\ns\times\ns$ matrix partitions, $U'_{(ij)}$, are diagonal, with
diagonal elements the diagonals of $\Udes^\dag\Usb_{(ij)}$.  Thus,
$\Gam_{ij}=\trace\ \Udes^\dag\Usb_{(ij)}=\trace\ U'_{(ij)}$. Now
replace the trace operation by the vector products, that is,
$\Gam_{ij}=w_i^T U' w_j$ where $w_i$ is an $\ns\nb\times 1$ vector
with $\ns$ ones in elements $1+(i-1)\ns$ to $i\ns$ and zeros
elsewhere. Hence we can write $\Gam=W^T U' W$ where $W=[w_1\ \cdots\
w_\nb]$ is $\ns\nb\times\nb$. Observe that $\norm{W}_2=\sqrt{\ns}$ and
since $U'$ is constructed from elements of unitary matrices $\Udes$
and $\Usb$, then $\norm{U'}_2\leq 1$. It therefore follows that
$\norm{\Gam}_2=\norm{W^TU'W}_2\leq
\norm{W}_2^2\norm{U'}_2\leq\ns$. This completes the proof.

\equote

%%%%%%%%%%%%%%%%%%%%%%%%%%%%%%%%
\newpage
\bc\subsection*{Acknowledgements}\ec

This work has been funded under the DARPA QuIST Program (Quantum
Information Science \& Technology).  R. Kosut would like to thank
Abbas Emami-Naeini (SC Solutions), Paul van Dooren (Catholic
University of Louvain), and Oliver Kosut (Cornell University) for help
with the proof of \refeq{fiddist01}.

%%%%%%%%%%%%%%%%%%%%
%\bibliographystyle{plain}
%\bibliographystyle{unsrt}
\bibliographystyle{alpha} 
\bibliography{D:/robert/tex/rlk}
%%%%%%%%%%%%%%%%%%%%%%%%%%%%%%%%%%%%%%%%%

\end{document}